\documentclass[twocolumn,showpacs,preprintnumbers,amsmath,amssymb,superscriptaddress,aps,prl]{revtex4}

\usepackage{graphicx}

\begin{document}

\title{A new determination of the fine structure constant based on
Bloch oscillations of ultracold atoms in a vertical optical lattice}

\author{Pierre~Clad\'e}
\affiliation{Laboratoire Kastler Brossel, Ecole Normale
Sup\'erieure, CNRS, UPMC, 4 place Jussieu, 75252 Paris Cedex 05,
France }
\author{Estefania~de~Mirandes}
\affiliation{Laboratoire Kastler Brossel, Ecole Normale
Sup\'erieure, CNRS, UPMC, 4 place Jussieu, 75252 Paris Cedex 05,
France }
\author{Malo~Cadoret}
\affiliation{Laboratoire Kastler Brossel, Ecole Normale
Sup\'erieure, CNRS, UPMC, 4 place Jussieu, 75252 Paris Cedex 05,
France }
\author{Sa\"\i da~Guellati-Kh\'elifa}
\affiliation{INM, Conservatoire National des Arts et M\'etiers,
292 rue Saint Martin, 75141 Paris Cedex 03, France}
\author{Catherine~Schwob}
\affiliation{Laboratoire Kastler Brossel, Ecole Normale
Sup\'erieure, CNRS, UPMC, 4 place Jussieu, 75252 Paris Cedex 05,
France }
\author{Fran\c cois~Nez}
\affiliation{Laboratoire Kastler Brossel, Ecole Normale
Sup\'erieure, CNRS, UPMC, 4 place Jussieu, 75252 Paris Cedex 05,
France }
\author{Lucile~Julien}
\affiliation{Laboratoire Kastler Brossel, Ecole Normale
Sup\'erieure, CNRS, UPMC, 4 place Jussieu, 75252 Paris Cedex 05,
France }
\author{Fran\c cois~Biraben}
\affiliation{Laboratoire Kastler Brossel, Ecole Normale
Sup\'erieure, CNRS, UPMC, 4 place Jussieu, 75252 Paris Cedex 05,
France }

\begin{abstract}
We report an accurate measurement of the recoil velocity of
$^{87}Rb$ atoms based on Bloch oscillations in a vertical
accelerated optical lattice. We transfer about $900$ recoil
momenta with an efficiency of $99.97\%$ per recoil. A set of $72$
measurements of the recoil velocity, each one with a relative
uncertainty of about $33$~ppb in $20$~min integration time, leads
to a determination of the fine structure constant $\alpha$ with a
statistical relative uncertainty of $4.4$~ppb. The detailed
analysis of the different systematic errors yields to a relative
uncertainty of $6.7$~ppb. The deduced value of $\alpha^{-1}$ is
$137.03599878(91)$.
\end{abstract}

\maketitle

The fine structure constant $\alpha$ plays an important role among
all the physical constants because it sets the scale of
electromagnetic interactions. Therefore, it can be measured in
different fields of physics and so be used to test the consistency
of the physics. In the Codata adjustment \cite{codata02}, all
accurate known determinations of $\alpha$ are used to give the
best estimate of $\alpha$ (labelled $\alpha_{2002}$ for 2002
adjustment). But as pointed out in \cite{codata02}, the actual
estimate $\alpha_{2002}$ is only determined by two data and in
fact mainly by the electron magnetic moment anomaly $a_{e}$
experiment. This lack of redundancy in input data is a key
weakness of the Codata adjustment. For example $\alpha_{2002}$
differs from $\alpha_{1998}$ by more than one sigma mainly because
of some revisions in the complicated theoretical expression of
$a_{e}$ from which $\alpha$ is deduced \cite{codata02}. Accurate
determinations of $\alpha$ by completely different methods are
absolutely needed. A competitive determination of $\alpha$ with
respect to the $a_{e}$ experiment is actually the measurement of
the ratio $h/m_{Cs}$ (where $h$ is the Planck constant and
$m_{Cs}$ is the mass of the Cesium atom) using ultracold atom
interferometry \cite{wicht}. The fine structure constant is
related to the ratio $h/m_X$ by \cite{Taylor}:

\begin{equation}
       \alpha^2=\frac{2R_\infty}{c}\frac{A_r(X)}{A_r(e)}\frac{h}{m_X}
\label{eqn1}
\end{equation}
where several terms are known with a very small uncertainty: $8
\times 10^{-12}$ for the Rydberg constant $R_\infty$ \cite{Schwob,
Udem} and $4.4\times10^{-10}$ for the electron relative mass
$A_r(e)$ \cite{codata02}. The relative atomic mass of X is known
with relative uncertainty less than $2.0\times10^{-10}$ for Cs and
Rb atoms \cite{Bradley}.

\begin{figure}
\includegraphics{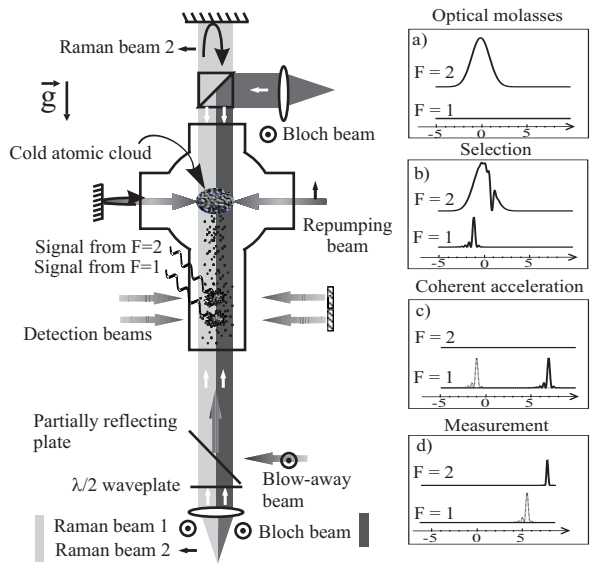}
\caption{\label{fig:1}Left. Experimental setup: the cold atomic
cloud is produced in a MOT (the cooling laser beams are not
shown), the Raman and the Bloch beams are in vertical geometry and
the detection zone is at 15 cm below the MOT. Right. Evolution of
the velocity distribution (in $v_r$ unit) during one experimental
cycle providing one point in the final velocity distribution shown
in Fig.\ref{fig:3} (see the text and \cite{Battesti}).}
\end{figure}

In this letter, we report a new determination of the fine
structure constant $\alpha$ deduced from the measurement of the
ratio $h/m_{Rb}$ based on Bloch oscillations. We describe a
sophisticated experimental method to measure accurately the recoil
velocity of a Rubidium atom when it absorbs or emits a photon. The
principle of this experiment is already described in a previous
paper \cite{Battesti}: by using velocity-selective Raman
transitions, we measure the variation of the atomic velocity
induced by a frequency-chirped standing wave. This coherent
acceleration arises from a succession of stimulated Raman
transitions where each Raman transition modifies the atomic
momentum by $2 \hbar k$ ($k = 2 \pi/\lambda$, $\lambda$ is the
laser wavelength), leaving the internal state unchanged. The
acceleration process can also be interpreted in terms of Bloch
oscillations in the fundamental energy band of an optical lattice
created by the standing wave \cite{Ben Dahan}: the atomic momentum
evolves by steps of $2\hbar k$, each one corresponding to a Bloch
oscillation. After $N$ oscillations, we release adiabatically the
optical lattice and we measure the final velocity distribution
which corresponds to the initial one shifted by $2Nv_r$
($v_r=\hbar k/m$ is the recoil velocity).
 In comparison with our prior setup
\cite{Battesti}, the Bloch beams (optical lattice) and the Raman
beams (velocity measurement) are now in vertical geometry
(Fig.\ref{fig:1}.Left).
This scheme is more suitable to achieve a high precision
measurement of the recoil velocity, because it allows us to
increase significantly the number of transferred momenta.

An atomic sample of $3\times 10^7$  atoms ($^{87}$Rb) is produced
in a magneto-optical trap (MOT), followed by a $\sigma^+ -
\sigma^-$ optical molasses. The final temperature of the cloud is
$3 \mu$K, its radius at $1/\sqrt{e}$ is $\sim 600 \mu$m and all
the atoms are in the hyperfine state $F=2$. An optical Zeeman
repumper (resonant with the $F=2$, $F^\prime=2$ transition)
transfers the atoms to the $F=2, m_F=0$ hyperfine state.

\begin{figure}
\includegraphics[width=.3\textwidth]{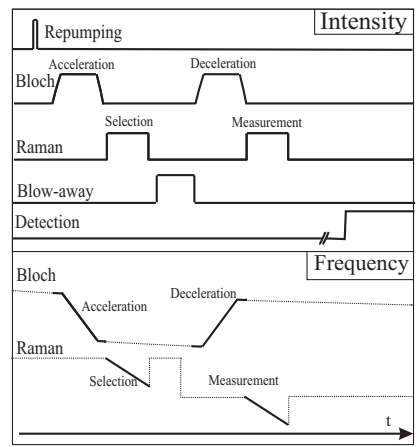}
\caption{\label{fig:2} Intensity and frequency timing of the
different laser beams for the acceleration-deceleration sequence.
(The scale of the frequency is not the same for the Bloch and the
Raman beams).}
\end{figure}

Then, a narrow velocity class is selected to $F=1~,m_F=0$ by using
a counter-propagating Raman $\pi$-pulse. The non selected atoms
are blown away using a resonant laser beam. After the acceleration
process described later, the atomic velocity distribution is
probed using a second Raman $\pi$-pulse from F=1 to F=2
(Fig.\ref{fig:1}.Right). The population in both levels is detected
using a time of flight technique \cite{Clairon} . The Raman beams
are produced by two stabilized laser diodes. Their beat frequency
is precisely controlled by a frequency chain allowing to easily
switch the Raman frequency detuning from the selection
($\delta_{sel}$) to the measurement ($\delta_{mes}$). One of the
lasers is stabilized on a highly stable Fabry-Perot cavity and its
frequency is measured by counting the beatnote with a two-photon
Rb standard \cite{Touari}. The frequency of one Raman beam is
linearly swept in order to compensate the Doppler shift induced by
the fall of the atoms (Fig.\ref{fig:2}) (with the same slope for
the selection and the measurement). The Raman beams power is
$8$~mW and their waist is 2~mm. To reduce photon scattering and
light shifts, they are blue detuned by 1~THz from the D2 line. The
duration of the $\pi$ pulse is $3.4$~ms: thus, the width of the
selected velocity class is $v_r/50$. In order to reduce the phase
noise, the Raman beams follow the same optical path: they come out
from the same fiber and one of them is retroreflected
(Fig.\ref{fig:1}.Left).

\textit{Coherent acceleration}. As shown in our previous work
\cite{Battesti}, Bloch oscillations of atoms in an optical lattice
are a very efficient tool to transfer a large number of recoil
momenta to the selected atoms in a short time. The optical lattice
results from the interference of two counter-propagating beams
generated by a Ti-Sapphire laser (waist of 2~mm), whose frequency
is stabilized on the same Fabry-Perot cavity used for the Raman
beams and is blue detuned by $\sim 40$~GHz from the one photon
transition. The optical lattice is adiabatically raised in
500~$\mu$s in order to load all the atoms into the first Bloch
band. To perform the coherent acceleration, the frequency
difference of the two beams is swept linearly within 3~ms using
acousto-optic modulators. Then, the lattice intensity is
adiabatically lowered in 500~$\mu$s to bring atoms back in a well
defined momentum state. The optical potential depth is $70~E_r$
($E_r=\hbar^2 k^2/2m$ is the recoil energy). With these parameters
the spontaneous emission is negligible. For an acceleration of
$2000$~ms$^{-2}$ we transfer 900 recoil momenta in 3~ms with an
efficiency  of $99.97\%$ per recoil. To prevent the atoms reaching
the upper windows of the vacuum chamber, we use a double
acceleration scheme (see Fig.\ref{fig:2}): instead of selecting
atoms at rest, we first accelerate them using Bloch oscillations
and then we perform the three steps sequence:
selection-acceleration-measurement. In this way the atomic
velocity at the measurement step is close to zero.

In the vertical direction, an accurate determination of the recoil
velocity would require a measurement of the gravity $g$. In order
to get rid of gravity, we make a differential measurement by
accelerating the atoms in opposite directions (up and down
trajectories) keeping the same delay between the selection and the
measurement $\pi$-pulses. The ratio $\hbar/m$ can then be deduced
from
\begin{equation}
\frac{\hbar}{m}= \frac{(\delta_{sel}-\delta_{meas})^{up} -
(\delta_{sel}-\delta_{meas})^{down}}{2(N^{up}+N^{down})k_B
(k_1+k_2)} \label{eq:2}
\end{equation}
where $(\delta_{meas}-\delta_{sel})^{up/down}$ corresponds
respectively to the center of the final velocity distribution for
the up and the down trajectories, $N^{up/down}$ are the number of
Bloch oscillations in both opposite directions, $k_B$ is the
wavevector of the Bloch beams and $k_1$ and $k_2$ are the
wavevectors of the Raman beams.
\begin{figure}
\includegraphics{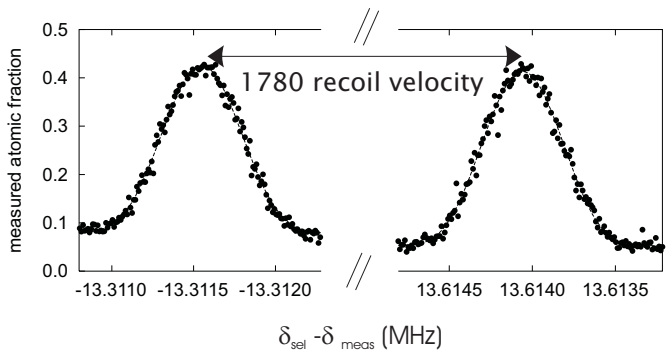}
\caption{\label{fig:3} Typical final velocity distribution for the
up and down trajectories.}
\end{figure}
In Fig.\ref{fig:3} we present two typical velocity distributions
for $N^{up} = 430$ and $N^{down} = 460$. The effective recoil
number is then $2 (N^{up}+N^{down})= 1780$. The center of each
spectrum is determined with an uncertainty of $1.7$~Hz ($\sim
v_r/10000$) for an integration time of $5$~min.

The  contribution of some systematic effects (energy levels
shifts) to $\delta_{sel}$ or $\delta_{meas}$ is inverted when the
directions of the Raman beams are exchanged. To improve the
experimental protocol, for each trajectory, the Raman beams
directions are reversed leading to the record of two velocity
spectra. When the atoms follow exactly the same up or down
trajectories, these systematic effects are cancelled by taking the
mean value of these two measurements. Finally one determination of
$\alpha$ is obtained from four velocity spectra ($20$~min of
integration time).

\begin{figure}
\includegraphics{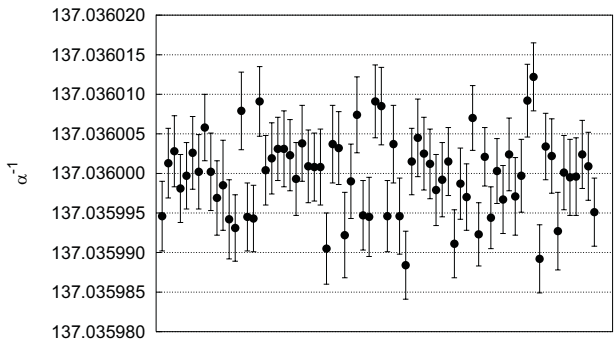}
\caption{\label{fig:4} Chronological display of $72$
determinations of $\alpha^{-1}$.}
\end{figure}

The Fig.\ref{fig:4} presents a set of $72$ determinations of the
fine structure constant $\alpha$. From the uncertainty of each
spectrum center we deduce the standard deviation of the mean. For
these $n=72$ measurements this relative uncertainty is $3.9$~ppb
with $\chi^2\simeq90$. Consequently, the resulting statistical
relative uncertainty on $\alpha$ is $3.9\times
\sqrt{\chi^2/(n-1)}=4.4$~ppb.


---\textit{Systematic effects analysis}--- We detail now all the
different systematic effects taken into account to determine the
final value of $\alpha^{-1}$ and its uncertainty.

\textit{Laser frequencies}: The frequency of the reference
Fabry-Perot cavity on which the Bloch and the Raman lasers are
stabilized is checked several times during a the 20~min
measurement, with respect to the Rb standard. The frequency drift
is 1~MHz and we deduce the mean laser frequency with an
uncertainty smaller than 100~kHz. Thus, we assume a conservative
uncertainty of 300~kHz for the absolute determination of the
different laser frequencies, which corresponds to 0.8~ppb on
$\alpha^{-1}$.

 \textit{Beams alignment}: We have measured the
fiber-fiber coupling of the counterpropagating Bloch and Raman
beams. It varies by less than 10\% with respect to the maximum
coupling. That corresponds to a maximum misalignment of
$3.1\times10^{-5}$~rad between the Raman beams and of
$1.6\times10^{-4}$~rad between the Bloch beams. The maximum
systematic effect on $\alpha^{-1}$ is of $-4\times10^{-9}$. Thus,
we correct $\alpha^{-1}$ by $(-2\pm 2)$~ppb.

\textit{Wavefront curvature and Gouy phase}: As the experimental
beams are not plane waves, we have to consider the phase gradient
in (\ref{eq:2}) instead of wavevectors $k$. For a Gaussian beam,
the phase gradient along the propagation axis is
\begin{equation}
    \frac{\mathrm{d}\phi}{\mathrm{d}z} = k - \frac2{kw^2(z)}
    - k \frac{r^2}{2R^2(z)} \frac{\mathrm{d}R}{\mathrm{d}z}
\label{eq:3}
\end{equation}
where $r$ is the radial distance from the propagation axis, $w(z)$
is the beam radius and $R(z)=z(1+(z_r/z)^2)$ is the curvature
radius. The first corrective term (Gouy phase) originates from the
spread on the transverse momenta, which is inversely proportional
to the beam transverse spatial confinement. The second term comes
from the spatial variation of the phase due to the curvature
radius. We have measured $w(z)$ and $R(z)$ with a wavefront
analyzer. The effective radial distance from the propagation axis
is determined by the size of the atomic cloud ($600 \mu$m) and a
possible misalignment of the Bloch beam with respect to the atomic
cloud. This misalignment is at maximum estimated at $500 \mu$m.
The correction to $\alpha^{-1}$ is $(-8.2\pm4)$~ppb. This is our
dominant systematic effect.

\textit{Magnetic field}: Residual magnetic field gradients
contribute to the systematics in two ways. Firstly there is a
second order Zeeman shift of the energy levels which induces an
error in the Raman velocity measurement. Secondly, the quadratic
magnetic force modifies the atomic motion between the selection
and the measurement. We have precisely measured  the spatial
magnetic field variations using copropagating Raman transitions.
The Zeeman level shift is not totally compensated by changing the
direction of the Raman beams because the two up (or down)
trajectories are not completely identical. They differ by about
300$~\mu$m, leading to a differential level shift of about
($0.3\pm 0.1$)~Hz and a $\alpha^{-1}$ correction of $(6.6\pm
2)$~ppb. The magnetic force changes the atomic velocity by
$(2.3\pm0.7)\times10^{-6}$ recoil velocity. We correct
$\alpha^{-1}$ by $(-1.3\pm0.4)$~ppb.

\textit{Gravity gradient}: Gravity is not totally compensated
between up and down trajectories because they differ by about
10~cm. The correction to $\alpha^{-1}$ is $(0.18\pm 0.02)$~ppb.

\textit{Light shifts}: In principle, light shifts are compensated
in three ways: between the selection and the measurement Raman
pulses, between the upward and downward trajectories and when the
Raman beams direction is changed. However this effect is not
totally cancelled. This is firstly due to a different intensity at
the selection and at the measurement because of the expansion of
the cloud, secondly to spatial intensity gradient along the beams,
and thirdly to intensity variations between the two Raman
configurations.  We calculate an effect of less than $\pm0.2$~ppb
on $\alpha^{-1}$. There is also a two-photon light shift due to
the copropagating Raman beams coming out from the same fiber
(before retroreflecting one of them). Its effect is larger at the
measurement when Raman beams are the closest to the copropagating
resonance and then corresponds to a correction on $\alpha^{-1}$ of
$(-0.5\pm0.2)$~ppb.

\textit{Index of refraction}: In a dispersive media of index $n$,
the laser wavelength $\lambda$ becomes $\lambda/n$ and then the
photon momentum transfer is $n\hbar k$. Recently, this change of
the atomic recoil momentum has been observed in a Kapitza-Dirac
interferometer \cite{Campbell}.
In our experiment, we have measured the total $^{87}$Rb +
$^{85}$Rb background vapor density as $8\times10^8$~at/cm$^3$. The
corresponding refractive index for the Bloch and Raman beams is
$(n-1) \simeq-7.2\times10^{-10}$ and $(n-1)\simeq-
3.6\times10^{-11}$ respectively. Thus we correct $\alpha^{-1}$ by
$(-0.37\pm0.3)$~ppb.
%
The initial density of the cold atoms is about
$1\times10^{10}$~at/cm$^3$, leading to a refractive index
$(n-1)_{\mathrm{sel}} \simeq -4\times10^{-10}$ at the selection.
The Bloch beams detuning is only 40~GHz. However, after the
selection, the atomic density is lower by at least a factor 50,
thus $(n-1)_{\mathrm{Bloch}} \simeq -2\times 10^{-10}$. Finally
for the Raman measurement $(n-1)_{\mathrm{meas}} \simeq
-10^{-12}$. We emphasize that the effect of the refractive index
of the cold cloud is different than the effect of the background
vapor refractive index. Especially, we have to take into account
the motion of the dispersive medium (cold cloud) in the global
momentum conservation and in the Doppler effect.
Indeed, in the case of the Bloch beams, the accelerated cloud is
itself the dispersive medium. If one has a 100\% transfer
efficiency, momentum conservation seems to indicate that the
refractive index of the cloud does not modify the recoil momentum
transferred to the atoms. In our experiment, we have a 99.95\%
efficiency per Bloch oscillation, which would correspond to a
modification of the atomic recoil momentum of about
$5\times10^{-4}\times(n-1)$, leading to a negligible effect. Let
us now consider the Doppler effect of an atom moving at the
average velocity of the cloud, during a Raman pulse: in the frame
of the cloud the length of the optical path is constant with time,
so the Doppler effect is independent from the refractive index of
the cloud. However, there is a small effect due to the atomic
recoil. It would lead to an effect on the recoil measurement of
the order of $(n-1)/N$ (where $N$ is the number of Bloch
oscillations), which is also negligible. A more detailed analysis
is in \cite{phD}. Nerveless, we have adopted for the refractive
index effect due to the cold cloud a conservative uncertainty of
$3\times 10^{-10}$ on $\alpha^{-1}$.\\ In the table
\ref{tab:table1}, we summarize the different systematic effects on
$\alpha^{-1}$. Our determination of $\alpha^{-1}$ is
$137.03599878(91)~[6.7\times 10^{-9}]$. This value is in good
agreement with the two competitive determinations based on atom
interferometry
$\alpha^{-1}(Cs)=137.0360001(11)~[7.7\times10^{-9}]$ and the $g-2$
experiment $\alpha^{-1}(a_e)=137.03599880(52)~[3.8\times10^{-9}]$
\cite{codata02}.

In conclusion, we have developed a powerful experimental approach
to measure accurately the atomic recoil velocity. Thanks to the
high efficiency of Bloch oscillations ($>99.97\%$ per recoil), we
are able to transfer $900$ photon momenta. To our knowledge, this
is the highest number of recoils ever transferred coherently to
any physical system. Our non interferometric measurement achieves
a precision comparable to the best interferometric measurement
\cite{wicht}. An even more rigorous control of some systematics
will be undertaken to reduce the uncertainty on a future
determination of $\alpha$.

\begin{table}
\caption{\label{tab:table1} Error budget (relative uncertainty in
ppb). }
\begin{ruledtabular}
\begin{tabular}{lcr}
\multicolumn{1}{l}{Source}
&Relative uncertainty \\
\hline Laser frequencies&0.8\\
Beams alignment& 2\\
Wavefront curvature and Gouy phase & 4\\
2nd order Zeeman effect & 2 \\
Quadratic magnetic force & 0.4\\
Gravity gradient& $0.02$ \\
light shift (one photon transition) & 0.2\\
light shift (two photon transition) & 0.2 \\
Index of refraction cold atomic cloud&0.3 \\
Index of refraction background vapor& 0.3 \\ \hline \hline
Global systematic effects& 5.0\\
Statistical uncertainty& 4.4\\
\end{tabular}
\end{ruledtabular}
\end{table}

\begin{acknowledgments}
This experiment is supported in part by the Bureau National de
M\'etrologie (contract 033006) and by the R\'egion Ile de France
(contract SESAME E1220). The work of E.~de~Mirandes is in keeping
with joint Ph.D thesis between the LENS,
($Universit\acute{a}~di~Firenze$) and Universit\'e Pierre et Marie
Curie and is supported by European community
(MEST-CT-2004-503847).
\end{acknowledgments}

\end{document}